\documentclass[twocolumn,aps,prb,superscriptaddress,10pt]{revtex4}
\usepackage{graphicx}
\usepackage{amsmath}
\usepackage{txfonts}
\usepackage{amsfonts}
\usepackage{amssymb}

\begin{document}

\title{Spin polarized Charge Trapping and Transfer at a HgTe Topological Insulator Quantum
Dot}
\author{L. Z. Lin}
\affiliation{SKLSM, Institute of Semiconductors, Chinese Academy of Sciences, Beijing,
China}
\author{Zhenhua Wu}
\affiliation{Key Laboratory of Microelectronics Devices and Integrated Technology,
Institute of Microelectronics, Chinese Academy of Sciences, Beijing, China}
\email{wuzhenhua@ime.ac.cn}

\begin{abstract}
This work presents theoretical demonstration of a carrier trap unit formed
by dual topological insulator constrictions (TIC) on the HgTe/CdTe quantum well
(QW) with inverted band structures. The sample of HgTe/CdTe QW is patterned
into a Hall bar device and a topological quantum dot is created by adding
split gate electrodes closely on the QW. In sharp contrast to conventional semiconductor
quantum dots, the presence or absence of topological edge states in the
proposed quantum hall bar system leads to distinct propagating/insulating
state of the TICs with large on/off ratio. This topological quantum dot
functions as a carrier trap memory element with near perfect program/erase
efficiency by proper adjusting the voltages applied to the split gates. For
completeness, we also demonstrate that the Rashba spin orbit interaction in
the quantum dot does not destroy the topological edge states and have
negligible impact on the conductance of the quantum hall bar. The rapid
oscillations in conductance can be suppressed when applying a perpendicular
magnetic field in the quantum dot.
\end{abstract}

\maketitle




\section{INTRODUCTION}

Manipulation of the transport properties in topological insulators (TI)\cite%
{Kane,Fu,Bernevig1,Hasan,Qi}, such as HgTe QWs \cite{Bernevig2,Yang,Konig}, $%
Bi_{1-x}Sb_{x}$\cite{Hsieh}, $Bi_{2}Se_{3}$\cite{Xia}, and $Bi_{2}Te_{3}$%
\cite{Chen}, have attracted rapidly growing interests over the past decade%
\cite{Hsieh2,Sonin,Cheng,Williams,Culcer,Lin,Lu,Hong,Siu,Bauer,Mani}. In two
dimension (2D), a topological insulator is also known as the quantum spin
Hall (QSH)insulator, with insulating bulk states and gapless helical edge
states. The QSH insulator state was predicted to occur in HgTe QWs with
inverted band structure\cite{Bernevig2,Yang} and was experimentally
demonstrated.\cite{Konig} These gap-less helical edge states are robust
against non-magnetic impurities, crystalline defects, and distortion of the
surface, et.al, due to the protection of time-reversal symmetry and non-zero
topological $\mathbb{Z}_{2}$ invariant.\cite{Kane,Sheng,Xu,Wu,Fu2} These
robust helical edge states provide new functional units for the construction
of future nano-electronic devices.

The observation of conductance quantization of a nano-scaled constriction in
2DEG, the so-called quantum point contact (QPC), was a hallmark of
mesoscopic transport physics.\cite{Wees,Wharam} The constriction can be
created by etching or adding split gate electrodes on top of the sample.
Since then, constrictions in various systems have been studied extensively
in various systems. Such constrictions are important building blocks of
quantum dots and qubits for many potential device applications e.g.,
single-electron transistors, quantum computers.\cite%
{Loss,Imamoglu,Hanson,Kastner,Galindo} After the recent discovery of the
aforementioned new state of quantum matter, i.e., TI, the topological
insulator constrictions in QSH system are of particular interest and
offering an elegant way to control the spin and charge transport. Currently,
most works focus on the properties of a single TIC under different
modulation mechanisms \cite{Zhou,Teo,Strom,Krueckl,Zhang,Huang,Klinovaja}.
It is highly desirable to study the impact of dual/multiple TICs
configurations or more complex hybrid structures on the modulation, so as to
achieve more functionable electric control of the carrier transport related
with the topological helical edge states.

In this work, we investigate theoretically the carrier transport properties
in experimentally realizable QSH system. We propose a topological quantum
dot formed in between dual TICs in HgTe/CdTe QW with inverted band
structure. The TICs are created by proper shaped split gate electrodes on
top of the QW. The applied voltages determine the extension of the depletion
regions underneath the split gates, making it possible to effectively tune
the width of each TIC independently. Rich transmission features will show up
with changing the widths of the TICs or the Fermi energy. A minigap can be
openned by the finite size effect at TICs, where the edge modes get coupled,
results in a conductance dip. This feature enables us to switch on/off the
edge channel by changing TIC widths, i.e., tuning the split gate voltages.
This configuration of a quantum dot isolated by dual TICs can be used as a
basic element in electronic storage devices. Utilizing robust topological
edge channels as well as a distinct switch between the QSH edge state and
bulk insulator state of TICs, give rise to superior device reliability,
endurance, power consumption in the new proposed devices. And bellow context
will illustrate these benefits in more detail. We also observed Fano-like or
Fabry-P\'{e}rot-like resonances arising from the interference in the
quantum dot between two TICS. For the additional reference of non-electric
modulation, the RSOI modulation considered in this work has a minor
influence on the conductance. When applied a perpendicular magnetic field in
the quantum dot, more conductance dips come out in the system, and the
backscattering process plays a crucial role at high magnetic field regime.

The paper is organized as follows. In Sec. II, we present the theory of
electron tunneling through dual TICs structures. In Sec. III, we show the
transmission, conductance, charge distribution of QSH bar system according
to charge trapping and propagation states to illustrate the role of edge
states in electronic storage devices. Finally, we give a brief conclusion in
Sec. IV.

\section{THEORETICAL MODEL}

\begin{figure}[tbp]
\centering
\begin{minipage}[c]{0.98\columnwidth}
\centering
\includegraphics[width=\columnwidth]{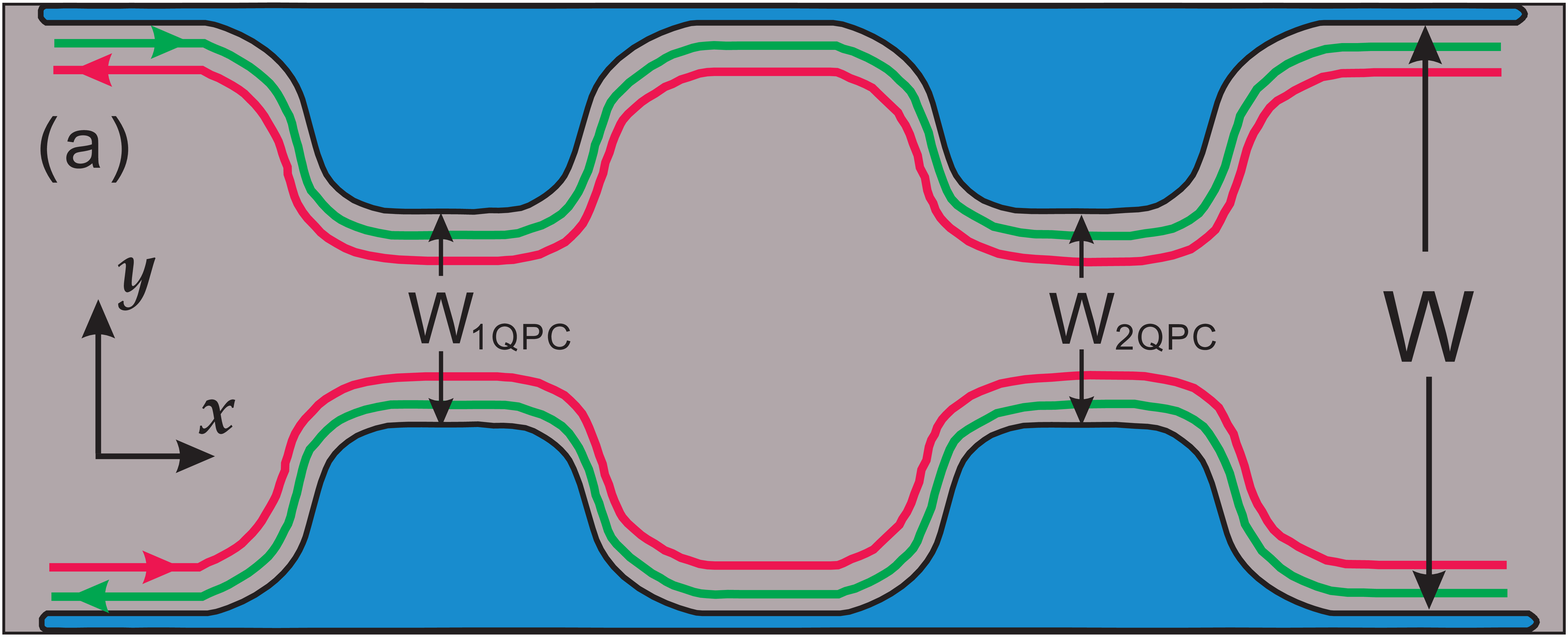}
\end{minipage}
\begin{minipage}[c]{1.0\columnwidth}
\centering
\includegraphics[width=\columnwidth]{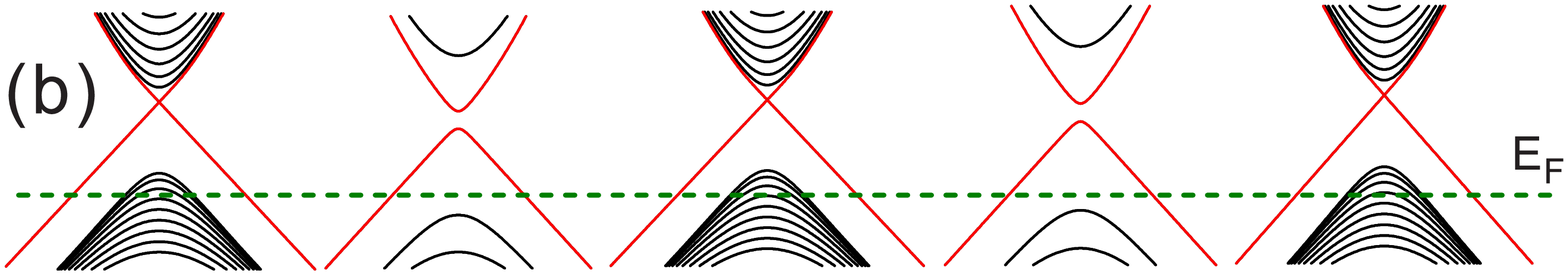}
\end{minipage}
\caption{(color online) (a) Schematic illustration of a topological quantum dot formed by dual TICs in a HgTe QSH
bar with inverted band structures. ${W_{TIC1}}$ (${W_{TIC2}}$) is the width
of the first (second) TIC. (b) The energy dispersions in the different
regions of the system. The green dashed line indicates the position of the
Fermi energy $E_F$. }
\label{fig:model}
\end{figure}

We consider a quantum dot with two TICs formed in a HgTe QSH bar, as shown
schematically in Fig.~\ref{fig:model}(a). Desired inverted band structures of HgTe QW in
the low-energy regime are guaranteed by BHZ model, i.e., a four-band
Hamiltonian obtained from the eight-band Kane model by neglecting the
light-hole and spin-split bands.[] The size confinement of the QW is treated
by a hard-wall potential $V$, $V(x,y)=V_{ext}$ for $y\in \lbrack {Y_{\min }}%
(x),{Y_{\max }}(x)]$ and $V(x,y)=\infty $ for $y<{Y_{\min }}(x)$ or $y>{%
Y_{\max }}(x)$, where ${Y_{\min }}(x)$ and ${Y_{\max }}(x)$ denote the lower
and upper boundaries of the QSH bar. Allowing the presence of an
perpendicular magnetic field, Zeem energy splitting ${g}_{{E}}{\mu }_{{B}}%
\mathbf{\sigma \cdot B}$ and vector potential $\mathbf{A}$ are induced. The
Landau gauge, $\mathbf{A}$ = $({A_{x}},0,0)$ is adopted in our calculation.
Finally with including the RSOI, such a single-particle under the modulation
of electric potential and magnetic field can be described by a 4 X 4
Hamiltonian in the basis $\left\vert {e\uparrow }\right\rangle ,\left\vert {%
hh\uparrow }\right\rangle ,\left\vert {e\downarrow }\right\rangle
,\left\vert {hh\downarrow \downarrow }\right\rangle $,

\begin{equation}
\begin{array}{l}
H=H_{0}+V+H_{Z} \\
=\left( {%
\begin{array}{cccc}
{{\varepsilon _{k}}+M(k)} & {A{k_{-}}} & {i\alpha {k_{-}}} & 0 \\
{A{k_{+}}} & {{\varepsilon _{k}}-M(k)} & 0 & 0 \\
{-i\alpha {k_{+}}} & 0 & {{\varepsilon _{k}}+M(k)} & {-a{k_{+}}} \\
0 & 0 & {-A{k_{-}}} & {{\varepsilon _{k}}-M(k)}%
\end{array}%
}\right)  \\
+\left( {%
\begin{array}{cccc}
V+{g}_{{E}}{\mu }_{{B}}{B}_{{z}} &  &  & 0 \\
& V+{g}_{{HH}}{\mu }_{{B}}{B}_{{z}} & 0 & 0 \\
& 0 & V+{g}_{{E}}{\mu }_{{B}}{B}_{{z}} &  \\
& 0 &  & V+{g}_{{HH}}{\mu }_{{B}}{B}_{{z}}%
\end{array}%
}\right) ,%
\end{array}%
\end{equation}%
where $k=({\pi _{x}},{k_{y}})$ is the in-plane momentum of electrons, ${\pi
_{x}=k}_{{x}}{+}\frac{e}{\hbar }A_{x}$, ${\varepsilon _{k}}=C-D(\pi
_{x}^{2}+k_{y}^{2})$, $M(k)=M-B(\pi _{x}^{2}+k_{y}^{2})$, ${k_{\pm }}={\pi
_{x}}\pm i{k_{y}}$, $\alpha $ is the RSOI strength, and $A,B,C,D$ and $M$
are the parameters describing the band structure of the HgTe/CdTe QW. Note
that the QSH state and band insulator state are characterized by the sign of
the parameter $M$, which is determined by the thickness of the HgTe/CdTe QW.%
\cite{Bernevig,Konig} For negative (positive) M, the QW is in QSH (bulk
insulator) state, respectively. For a quasi-one-dimensional (Q1D) QSH bar
system shown in Fig.~\ref{fig:model}(a), the transport property can be obtained by
discretizing the Q1D system into a series of transverse strips along the
transport direction with sharply narrowed widths at TICs controlled by the
split-gate modulation. By means of scattering matrix theory,\cite{Zhang2}
the transmission amplitude from the m-th input mode to n-th output mode ${%
t_{nR;mL}}$ are obtained. We can get the total conductance of the QSH system
at zero temperature from the Landauer-B\"{u}ttiker formula,\cite{Buttiker}
\begin{equation}
G^{0}(E_{F})={G_{0}}{\sum\limits_{m\in L,n\in R^{\prime }}}\frac{\nu _{n}^{R}%
}{\nu _{m}^{L}}{{\left\vert {{t_{nR;m^{\prime }L}(E}}_{{F}}{)}\right\vert }%
^{2}},
\end{equation}%
where the summation is over all propagation modes in the input/output leads
which are to the left/right of the quantum dot, ${G_{0}}={e^{2}}/h$ is the
conductance unit. At a finite temperature T, the ballistic conductance can
be written as
\begin{equation}
G(E_{F})=\int (-\frac{\partial f(E)}{\partial E})G^{0}(E_{F})dE,
\end{equation}%
where $f(E)=\{1+\exp [(E-E_{F})/k_{B}T]\}^{-1}$ is the Fermi-Dirac
distribution.

\section{RESULTS AND DISCUSSION}

\begin{figure}[tbhp]
\centering
\includegraphics [width=0.9\columnwidth]{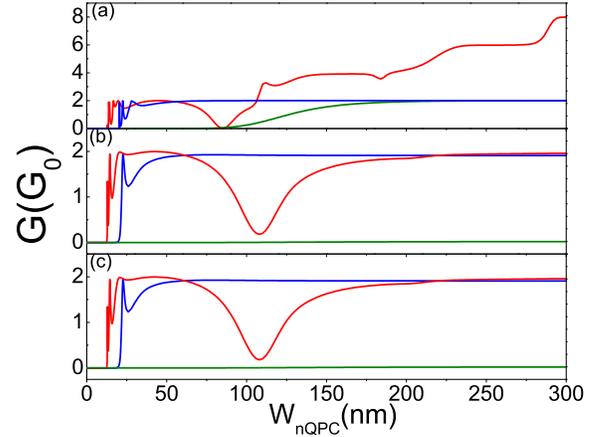}
\caption{(color online) The conductance as a function of the width of the
TICs (${W_{nQPC}}$ $n=1$ or 2) for tree different Fermi energies ${E_F} =$ $- 15$
meV (red lines), $10$meV (green lines), $-5$ meV (blue lines). (a) The width
of two QPCs are changed at the same time $\left( {{W_{nQPC}} = {W_{TIC1}} = {%
W_{TIC2}}} \right)$. (b)Fix the width of the second TIC (${W_{TIC2}=50}$ nm),
change the width of the first TIC (${W_{nQPC}}={W_{TIC1}}$). (c) Fix the
width of first TIC (${W_{TIC1}=50}$ nm), change the width of the second TIC (%
${W_{nQPC}}={W_{TIC2}}$). If it is not specified, the width of the QSH bar is always $W=300$ nm.}
\label{fig:G}
\end{figure}

We start by investigating the carrier transport properties in the QSH bar
system without RSOI nor magnetic field. Fig.~\ref{fig:G} shows the conductance
variations with the dual TICs width modulations in three different ways. Due
to the finite size effect,\cite{Zhang,Klinovaja} a finite gap can be opened
in the energy spectra [see Fig.~\ref{fig:model}(b)] and the gap keeps increasing when
reducing the width of TIC. If the Fermi energy locates in this opened energy
gap, there is no available propagating mode in the TIC region. Instead,
carriers move in evanescent modes that decay rapidly in the TIC region,
leading to the suppression of transmission. For extreme narrow TICs, we can
always observe vanished conductance as shown in Fig.~\ref{fig:G}(a) regardless the
incident Fermi energy we choose, since the gap is very large. For the green
curve, i.e., ${E_F} = 10$ meV, the Fermi energy is close to the Dirac point
of the QSH TI state. Even though only a small gap is opened when the width
of TICs is just bellow 100 nm, the Fermi energy has already located in this
gap in consistent with the conductance drop. For the blue curve, i.e., ${E_F}
= -5$ meV, the Fermi energy lies far from the Dirac point of the QSH TI
state, but still does not touch the edge of subbands (bulk propagating
modes) even when TICs are as wide as the leads. Therefore, we need much
larger gap in the TICs region with further reduced width to block the edge
channel. In the above two cases, only the edge state contributes to the
transport and thus the conductance saturates to the value of 2G0 as we relax
the TICs. Note that, when the incident energy is fixed to ${E_F} = - 15$
meV, the Fermi level is so apart from the Dirac point that it is able to
cross several bulk propagating subbands as the widths of TICs get larger
[see Fig.~\ref{fig:model}(b)]. As we expect, the conductance does not saturate with the
width of TICs but exhibits a step-like feature, which corresponds to the
opening new bulk propagating modes in TICs regions as they become wider.
Furthermore we can observe a dip in conductance curve around $W \approx 85$
nm. The conducting channels come from both the QSH TI edge states and the
bulk propagating states, the wave functions of the QSH TI states are
squeezed in the central quantum dot region and couple with the bulk states
in the narrow TICs. Due to the destructive quantum interference when the
charge carriers transmit or reflect at TICs, carriers are localized in the
quantum dot without transmission. In Fig.~\ref{fig:G}(b) and (c) we confirm that each
single TIC can effectively block the carrier transport, making our proposed
TI quantum dot be possible to work as a electronic storage device as we will
discuss in more detail later.

\begin{figure}[tbhp]
\centering
\includegraphics [width=0.9\columnwidth]{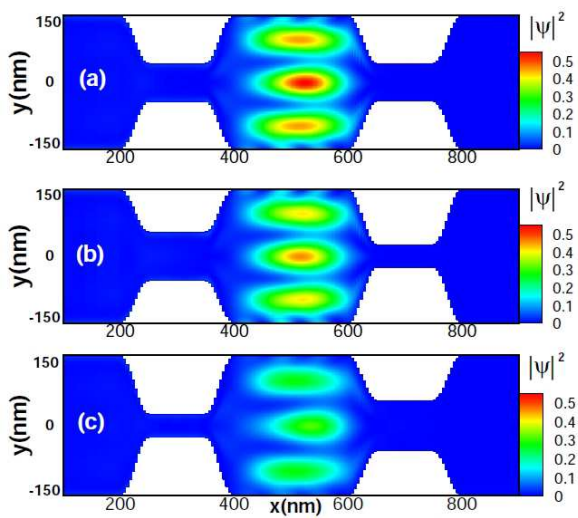}
\caption{(color online) The density distribution of the carrier in both the
edge states and bulk states. The Fermi energy is fixed at ${E_F} = - 15$ meV. The widths of TICs are as follows: (a) $%
{W_{TIC1}} = {W_{TIC2}} = 85$ nm. (b) ${W_{TIC1}} =108$ nm, ${W_{TIC2}} = 50$
nm. (c) ${W_{TIC1}} =50$ nm, ${W_{TIC2}} = 108$ nm.}
\label{fig:DOS}
\end{figure}

To test the analytical prediction of carrier localization, we plot the
density distribution of both topological edge states and bulk states for
three cases in Figs.~\ref{fig:DOS}(a)-(c), corresponding to the dips in Figs.~\ref{fig:G}(a)-(c)
respectively. The Fermi energy is -15 meV, the widths of dual TICs are as
follows, 1) $W_{TIC1}=W_{TIC2}=85$nm, 2) $W_{TIC1}=108$nm, $W_{TIC2}=50$nm,
3) $W_{TIC1}=50$nm, $W_{TIC2}=108$nm. Our calculations demonstrate the
localized bound states in the quantum dot. Carriers can be confined in these coupling quantum states with appropriate TICs widths and Fermi energy.
 Despite the similarity of charge trapping phenomenon as compared to conventional quantum dot, the physical content is different in our system, i.e., the bound states arise from the coupling between the topological edge states and the the bulk states rather than the common quantum confinement due to potential
barriers. We can adiabatically change the Fermi energy to approach such condition of charge trapping. It provides the possibility of storing information in these
coupling quantum states. However, 1) it is subtle to broke the edge channel by coupling, 2) the transition from trapping phase to transfer phase is rather smooth, and 3) The new mechanism of bound states in the quantum dot does not give rise to some functional benefits.
For the purpose of making an electronic storage element, very sharp transition is requested to boost device performance, e.g., quick switch on/off, low standby power, et.al.

\begin{figure}[tbhp]
\centering
\includegraphics [width=0.9\columnwidth]{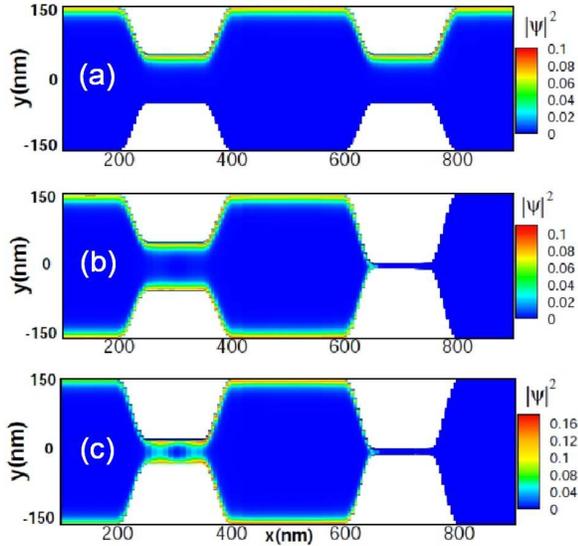}
\caption{(color online) The density distribution of the carrier in the edge
states. The Fermi energy is fixed at ${E_F} = - 5$ meV. The widths of TICs are as follows:
(a) ${W_{TIC1}} = {W_{TIC2}} = 100$ nm. (b) ${W_{TIC1}} =100$ nm, ${W_{TIC2}}
= 10$ nm. (c) ${W_{TIC1}} =50$ nm, ${W_{TIC2}}
= 10$ nm.  }
\label{fig:edgeDOS}
\end{figure}

To find possible solution, we consider the carrier transport purely in helical edge channels by tuning the Fermi energy to -5 meV. The bulk subband are too far from this Fermi energy to contribute to carrier transport.
In the absence of RSOI and magnetic field, we obtain a block diagonal Hamiltonian leading to the time-reversal symmetry for the upper and lower $2
\times 2$ blocks. Therefore we can calculate the carriers transport of spin-up or down state separately.
Fig.~\ref{fig:edgeDOS} show the density distribution of the helical edge states in the system for different TICs configurations. For simplicity, we only take the spin-up carriers in to consideration. In Fig.~\ref{fig:edgeDOS}(a), the TICs are adequate wide with remained edge channels in whole QSH bar device regions. The carriers propagate along the edge channel and perfectly transport through the topological quantum dot. If the second TIC is blocked as Fig.~\ref{fig:edgeDOS}(b). One can see clearly that the carriers from the left incident lead can transport along the top edge into the quantum dot and be reflected to the opposite edge with the
same spin orientation due to the helicity of the topological edge states. Then carriers leak out of the quantum dot easily also via the edge channels. Since the width of the first TIC is wide enough, the coupling between top and bottom helical edge stats is very weak. Carriers can transport freely in these edge channels that are well kept in the TICs. If we reduce the width of the first TIC as we did for the second TIC, the coupling between the edge states in the opposite edges become spectacular as we can see in Fig.~\ref{fig:edgeDOS}(c). The maximum local probability ${\left| {\psi
_{TES}^\sigma } \right|^2}(\sigma = \uparrow , \downarrow )$ is
approximately 0.17 between the dual TICs which is significantly enlarged
comparing to the configuration as shown in Fig.~\ref{fig:edgeDOS}(b), in which the maximum local probability ${\left| {%
\psi _{TES}^\sigma } \right|^2}(\sigma = \uparrow , \downarrow )$ is about
0.9]. This gap indicates that the carriers can't reflect back to left incident lead freely
and tend to be stored in the quantum dot. If we keep reducing the width of the TIC, the coupling is strong enough to break the topological edge modes and the first TIC is blocked as well as the second TIC. One can first apply the pinch-off voltage on
the split gate to block the second TIC, after the charges are injected from
left lead via the spin-polarized helical edge channel, then block the first TIC with the same technique. 

In sharp contrast to conventional semiconductor quantum dots, the quantum states in our topological quantum dot are fully spin-polarized helical edge states at a
appropriate Fermi energy in the bulk gap (in our discussion, ${E_F} = - 5 $ meV).
The trapped charges show ringlike density distributions near the boundary as we demonstrated in Fig.~\ref{fig:edgeDOS}(c). 
More importantly, the spin-angular momentum locking, guaranteed by the BHZ hamiltonian, gives rise to spin-polarized charge currents, i.e., the trapped spin-up carriers rotate clockwise and spin-down carriers rotate counterclockwise.~\cite{Chang}  If we inject spin polarized currents in to the quantum dot, the polarized spin can be stored as well as the charge in form of persistent spin/charge currents accounting for the topological protected edge states, which are robust against local perturbations. 
This feature sheds new light on constructing charge/spin trapping memory element for both electronic and spintronic devices.

\begin{figure}[tbhp]
\centering
\includegraphics [width=0.9\columnwidth]{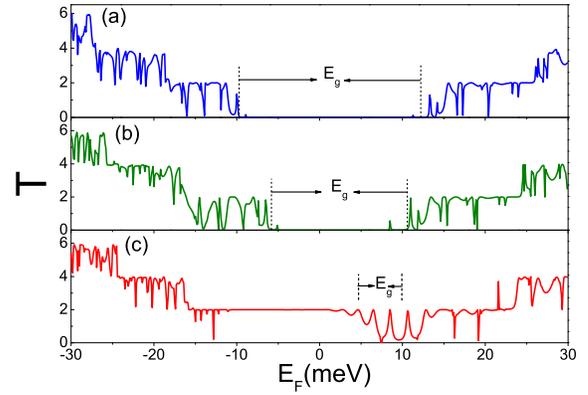}
\caption{(color online) Fermi energy dependence of the transmission T with fixed TICs width (${{W_{TIC1}} = {W_{TIC2}} = 100}$ nm) but for different band parameter M. (a) $M=8$ meV, (b) $M=4$
meV and (c) $M=-10$ meV. }
\label{fig:M2}
\end{figure}

To proceed further, we examine the impact of the HgTe QW thickness, which can be characterized by a band parameter M.
As we have discussed, M determines the presence of topological insulator (TI) edge state or only trivial bulk insulator (BI) state.~\cite{Bernevig2}
In Fig.~\ref{fig:M2}, we show the transmission probability $T$ as a function of
the Fermi energy ${E_F}$ with different band parameter $M$. The different
tunneling precess between BI $\left( {M> 0} \right)$ and TI $\left( {M < 0}
\right)$ can be observed when the widths of two TICs are fixed to 100 nm. 
For BI system, the transmission vanishes when the incident Fermi energy located in the gap of the TICs due to the absence of propagating modes.
For higher Fermi energy, transmission plateaus appear as well as many oscillations. Each plateau corresponds to a bulk subband in the TICs. The oscillations originate from
the  Fabry-P\'{e}rot interference between the transmitted and reflected electrons in the quantum dot.
For TI system, the resonances are observed even when the Fermi energy is located in the small TI edge gap.
When the incident Fermi energy located at a value in the bulk gap but not in the edge gap of the TICs, the transmission is nearly perfect accounting for the helical edge channels.
The oscillation of the transmission is smeared out since the backscattering
process is suppressed, i.e., few Fabry-P\'{e}rot mode are formed.
This observation provides us additional way to controlling the carrier transport properties in the proposed HgTe QSH bar device.

\begin{figure}[tbhp]
\centering
\includegraphics [width=0.9\columnwidth]{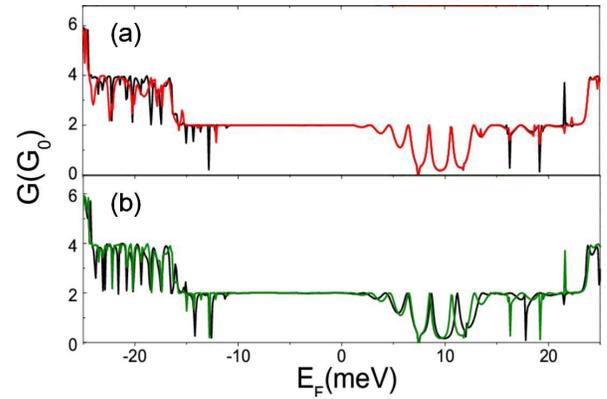}
\caption{(color online) The conductance as a function of the Fermi energy $%
E_F$ with including (a) RSOI $\protect\alpha=50$ meV nm (red lines) or (b) magnetic
field $B=1 T$ (green lines) in the quantum dot with fixed TICs width (${W_{1QPC}} = {W_{2QPC}} = 100$ nm).
The black lines indicates $\protect\alpha=0$ and $B=0$ in the quantum dot.}
\label{fig:RSOI}
\end{figure}

So far, we have studied the carrier transport through proposed QSH bar system
exclusively under the electric modulations. For completeness, we further investigate the carrier
transport through the HgTe QSH bar in the presence of RSOI or magnetic filed
in the quantum dot. Fig.~\ref{fig:RSOI}(a) shows the conductance as a function of the incident Fermi energy in the presence/absence of RSOI.
Since the topological edge states are immune to perturbation of RSOI, the conductance curves of different RSOI strength $\alpha=0$ and $\alpha=50$ meV are almost the same in the energy region $- 10$ meV $< {E_F} < $ $15$ meV. The conducting channels only come from topological edge states.
If the Fermi energy locates out of the bulk band gap, i.e., ${E_F} < $ $-10$
meV or ${E_F} >$ $15$ meV, the bulk modes are opened in the quantum dot. The RSOI would
induce a spin splitting of the bulk subbands, resulting in different Fermi
wave vectors ${k_\sigma }$ ($\sigma = \uparrow , \downarrow $) for spin-up and spin-down carriers. However the spin
splitting is quite small, so one can expect that this spin splitting only
leads to slight different behaviors comparing to the case without RSOI in the QD, which is clearly reflected in Fig.~\ref{fig:RSOI}(a).
Finally, we consider an external magnetic field applied perpendicularly to the
HgTe quantum dot. Fig.~\ref{fig:RSOI}(b) shows the conductance as a function of the incident
Fermi energy $E_F$ for $B=1$ T and $B=0$ T . Interestingly, we find that the positions of resonance conductance peaks
are obviously shifted in a finite magnetic filed of $B=1$ T. 
Because the magnetic field breaks the time-reversal symmetry and shifts the
Dirac point in the energy spectra.~\cite{Chang} Beyond the energy range with only topological edge state, the conductance behaviors are completely different when
we apply a finite magnetic field ($B=1 T$). More oscillations and sharp
dips are observed in our calculation. These dips indicate strong backscattering in the quantum dot. It can be understood by the semiclassical
picture: The magnetic
field can bend the trajectory of the carriers in favor of cyclotron motion. Charge carriers have more chance to meet with each other before transmit through the quantum dot. Therefore it will enhance the interference between different modes in the quantum dot. If the magnetic field is very large, cyclotron orbit radius is smaller than the size of the quantum dot.
The conductance can be fully suppressed. For small RSOI and magnetic field as we employed in above simulation, the topological edge states are well preserved. 
All the aforementioned features needed to make a spin-polarized charge trapping device element are kept in the proposed quantum dot even in presence of small RSOI and/or magnetic field.

\section{CONCLUSIONS}

In summary, we have theoretically investigated the carrier transport
through a topological quantum dot formed by dual TICs in HgTe QWs with inverted band structures. The
conductance of the system can be tuned by changing the width of the TICs or
the Fermi energy. we find that the transmission of carriers can be nearly perfect through the system even when Fermi energy is located in both the edge and bulk
gap, which would not happen in BI system. The transmission exhibits a
series of Fano resonances originating from the Fabry-P\'{e}rot modes localized in
the quantum dot between dual TICs by adjusting the Fermi energy ${E_F}$. with certain TICs configurations and proper incident Fermi energies, carriers can be trapped in the quantum dot in two different forms: 1) in the bulk bound states, 2) in the helical edge states.
Espatially for the second case, it may pave a new path of constructing charge/spin trapping memory element for both electronic and spintronic devices.
Finally we demonstrate that the proposed topological quantum dot memory element is immune to small RSOI or magnetic filed, as well as some other local perturbations.

\end{document}